\documentclass[%
 reprint,
 jmp,
superscriptaddress,
 amsmath,amssymb,
 aps,
]{revtex4-2}

\usepackage{braket}
\usepackage{graphicx}
\usepackage{dcolumn}
\usepackage{bm}

\usepackage[usenames,dvipsnames]{xcolor}
\usepackage[hyperindex,pdftex, breaklinks,colorlinks = true,linkcolor = blue,urlcolor=blue,citecolor=blue]{hyperref}

\begin{document}

\preprint{APS/123-QED}

\title{A reconfigurable silicon photonics chip for the generation of  frequency bin entangled qudits}

\author{Massimo Borghi}
\email{massimo.borghi@unipv.it}
\affiliation{Dipartimento di Fisica, Università di Pavia, Via Agostino Bassi 6, Pavia 27100, Italy}

\author{Noemi Tagliavacche}
\affiliation{Dipartimento di Fisica, Università di Pavia, Via Agostino Bassi 6, Pavia 27100, Italy}

\author{Federico Andrea Sabattoli}
\altaffiliation[Current address:]{ Advanced Fiber Resources Milan S.r.L, via Fellini 4, 20097 San Donato Milanese (MI), Italy.}
\affiliation{Dipartimento di Fisica, Università di Pavia, Via Agostino Bassi 6, Pavia 27100, Italy}

\author{Houssein El Dirani}
\altaffiliation[Current address:]{ LIGENTEC  SA,  224  Bd  John  Kennedy, 91100 Corbeil-Essonnes, France.}
\affiliation{Université Grenoble Alpes, CEA-Leti, Grenoble cedex 38054, France}

\author{Laurene Youssef}
\altaffiliation[Current address:]{ IRCER, Centre Européen de la Céramique, 12 rue Atlantis, 87068 Limoges, France.}
\affiliation{Univ. Grenoble Alpes, CNRS, CEA/LETI-Minatec, Grenoble INP, LTM, F-38054 Grenoble-France}

\author{Camille Petit-Etienne}
\affiliation{Univ. Grenoble Alpes, CNRS, CEA/LETI-Minatec, Grenoble INP, LTM, F-38054 Grenoble-France}

\author{Erwine Pargon}
\affiliation{Univ. Grenoble Alpes, CNRS, CEA/LETI-Minatec, Grenoble INP, LTM, F-38054 Grenoble-France}

\author{J.E. Sipe}
\affiliation{Department of Physics, University of Toronto, 60 St. George Street, Toronto, Ontario M5S 1A7, Canada}


\author{Marco Liscidini}
\affiliation{Dipartimento di Fisica, Università di Pavia, Via Agostino Bassi 6, Pavia 27100, Italy}

\author{Corrado Sciancalepore}
\altaffiliation[Current address:]{ SOITEC   SA,   Parc   Technologique   des Fontaines, Chemin des Franques, 38190 Bernin, France.}
\affiliation{Université Grenoble Alpes, CEA-Leti, Grenoble cedex 38054, France}

\author{Matteo Galli}
\email{matteo.galli@unipv.it}
\affiliation{Dipartimento di Fisica, Università di Pavia, Via Agostino Bassi 6, Pavia 27100, Italy}

\author{Daniele Bajoni}
\affiliation{Dipartimento di Ingegneria Industriale e dell’Informazione, Università di Pavia, Via Adolfo Ferrata 5, Pavia 27100, Italy 
}%

\date{\today}

\begin{abstract}
Quantum optical microcombs in integrated ring resonators generate entangled photon pairs over many spectral modes, and allow the preparation of high dimensional qudit states. Ideally, those sources should be programmable \textcolor{black}{and} have a high generation rate, 
\textcolor{black}{with} 
comb lines 
tightly spaced for the implementation of efficient qudit gates based on electro-optic frequency mixing. \textcolor{black}{While these} 
requirements cannot \textcolor{black}{all} be 
satisfied by a single 
\textcolor{black}{resonator device}, \textcolor{black}{for which there is a trade-off between high generation rate and tight bin  spacing, a promising strategy is} 
the use of multiple resonators, each generating photon pairs in specific frequency bins via spontaneous four-wave mixing. 

Based on this 
\textcolor{black}{approach} we present a programmable silicon photonics device for the generation of frequency bin entangled qudits, in which  bin spacing, qudit dimension, and bipartite quantum state can be reconfigured on-chip. Using resonators with a radius of $22\,\mu$m, we achieve a high brightness ($\sim \textrm{MHz}/(\textrm{mW})^2$) per comb line with a bin spacing of $15$ GHz, and fidelities above $85\%$  with maximally entangled Bell states up to a Hilbert space dimension of sixteen. By individually addressing each spectral mode, we realize states that can not be generated on-chip using a single resonator. We measure the correlation matrices of maximally entangled two-qubit and two-qutrit states on a set of mutually unbiased bases, finding fidelities exceeding $98\%$, and  indicating that the source can find application in high-dimensional secure communication protocols. 
\end{abstract}

\maketitle

\section{\label{sec:Introduction} Introduction}
Increasing the dimensionality of the Hilbert space of any quantum system 
is important in the development of large scale architectures for quantum technologies. This goal can be 
achieved \textcolor{black}{either} by increasing the number $n$ of parties involved in the quantum state\textcolor{black}{,} or their local dimension $d$.
\textcolor{black}{For photons, increasing $d$ presents less challenges than increasing $n$.} 
\textcolor{black}{Increasing $n$ leads to less tolerance to losses, and the lack of strong photon-photon interactions makes the realization of controlled gates very demanding \cite{knill2001scheme}. But multidimensional} 
\textcolor{black}{photon} states can be efficiently prepared and controlled \textcolor{black}{by employing one of the many photon} 
degrees of freedom 
\cite{erhard2020advances}, 
\textcolor{black}{including} orbital angular momentum \cite{zahidy2022photonic}, path  \cite{wang2018multidimensional}, time \cite{islam2017provably}, \textcolor{black}{and} frequency \cite{kues2017chip}, or 
\textcolor{black}{by using a} combination \textcolor{black}{of them} \cite{imany2019high,lu2020quantum}. 
\textcolor{black}{The qudits that can be produced} \textcolor{black}{are a greater resource than the qubits of two-dimensional systems, in that} 
\textcolor{black}{they can be exploited to demonstrate} 
stronger violations of 
Bell's inequalities \cite{collins2002bell}, \textcolor{black}{they can be used to} speed up 
 quantum algorithms \cite{wang2020qudits}\textcolor{black}{, and they can reduce} 
circuit depth for logic operations \cite{lanyon2009simplifying}. \textcolor{black}{Their use leads to} improved noise resilience in communication protocols \cite{cerf2002security}, and \textcolor{black}{they offer} the possibility 
\textcolor{black}{of} 
deterministic\textcolor{black}{ally} 
\textcolor{black}{controlling} operations between the multiple qubits that can be encoded in the Hilbert space of a single qudit \cite{vigliar2021error}. 

One of the strategies for the generation of high-dimensional states is to pump 
a parametric source of photon pairs \textcolor{black}{with a continuous-wave laser; this} 
generates time-frequency entanglement 
\cite{chang2021648,liu2019energy}. The large information content per biphoton can be exploited \textcolor{black}{either for} 
transmit\textcolor{black}{ting} a large alphabet in the time domain \cite{ali2007large,liu2019energy}, or for the deployment of wavelength division multiplexed networks for secure communication \cite{liu2020entanglement,joshi2020trusted}. The continuous spectra of photon pairs can be sliced into non-overlapping frequency bins to encode different logical states, an operation that 
\textcolor{black}{can} be \textcolor{black}{effected} 
by filtering the broadband emission with a Fabry-Perot cavity \cite{chang2021648}. Alternatively, one 
\textcolor{black}{can} engineer the 
 generation of photons in discrete spectral modes \cite{morrison2022frequency}. 
 
\textcolor{black}{On integrated platforms,} 
quantum optical  Kerr microcombs 
\textcolor{black}{generated by} ring resonators \textcolor{black}{are} 
important 
sources 
of high-dimensional frequency entanglement \cite{kues2017chip,imany201850,jaramillo2017persistent,lu2022bayesian}. Compared to bulk sources, quantum microcombs have distinctive properties, such as 
high brightness ($\sim \textrm{MHz}/(\textrm{mW})^2$), phase coherence over many spectral modes without active stabilization \cite{jaramillo2017persistent}, and a free spectral range (FSR) of tens of GHz \cite{imany201850} \textcolor{black}{that allows for the individual} 
address and 
\textcolor{black}{manipulation of} each spectral line \cite{kues2017chip}. In many applications, this is accomplished through a quantum 
\textcolor{black}{Fourier} processor (QFP), a tool for the coherent manipulation of frequency bin states  \cite{lu2018quantum}. It consists of a repeated sequence of electro-optic modulators (EOMs), used for the frequency mixing of the different modes, and waveshapers, 
\textcolor{black}{employed to} apply \textcolor{black}{a} different 
\textcolor{black}{phase shift to} 
each spectral line.  

A whole framework 
\textcolor{black}{of quantum information processing with frequency bins is being developed, leading to a new paradigm for} quantum  simulation and computation. 
Arbitrary local operations on qubits  have been demonstrated with near unit fidelity and success probability \cite{lu2020fully,lu2018electro,lu2019quantum}, and two-qubit gates have been 
proposed and experimentally realized with \textcolor{black}{both} heralded \cite{lukens2017frequency} and post selection schemes \cite{lu2019controlled}. A class of local qudit gates, which include\textcolor{black}{s} \textcolor{black}{those for} the discrete Fourier transform 
\cite{lu2022high} and the frequency hop operation \cite{lukens2019all}, 
\textcolor{black}{has} been demonstrated using the QFP \textcolor{black}{with} 
multi-tone driving of the EOMs. Promisingly, the number of optical elements in a QFP 
\textcolor{black}{that} is required to realize arbitrary qudit gates is predicted to linearly scale with the system size \cite{lukens2019all}.

Crucially, for 
\textcolor{black}{these} operations to be efficient the frequency bin spacing $\delta$ should not exceed the bandwidth of the EOMs, which is of the order of $\sim 50$ GHz for commercial devices.
If generation is to be done in a single microring resonator, the condition of closely spaced combs requires the use of a 
ring with a small FSR, and thus a radius of several hundred of microns. This   
\textcolor{black}{limits its} brightness, which scales with 
\textcolor{black}{the square of the FSR} \cite{liscidini2019scalable}. Recently, it has been shown that this trade-off can be 
\textcolor{black}{bypassed} with the use of multiple resonators, each associated with the generation of a pair of non-degenerate frequency bins \cite{liscidini2019scalable}, and this 
\textcolor{black}{was } demonstrated \textcolor{black}{experimentally} for a two qubit system \cite{clementi2023programmable,sabattoli2022silicon}. In \textcolor{black}{that} 
scheme, 
a single pump is resonant with all the rings\textcolor{black}{,} and triggers spontaneous four wave mixing (SFWM) in non-degenerate frequency bins\textcolor{black}{, which }
occurs because each ring has a different FSR
. \textcolor{black}{But this} requires the use of auxiliary resonators to ensure that the sources can be pumped \textcolor{black}{both} coherently 
\textcolor{black}{and} independently \cite{liscidini2019scalable}. 

In this work, we \textcolor{black}{generalize} 
this \textcolor{black}{strategy} 
\textcolor{black}{to} demonstrate the generation of programmable frequency-bin-entangled qudits on a silicon photonics chip. We use four identical rings, and offset their comb lines to create equally spaced frequency bins with a separation of $\delta$. This quantity can be reconfigured to allow flex-grid (in frequency) operation. Each ring  is pumped at a different wavelength with mutually coherent pumps. This configuration allows \textcolor{black}{us} to generate entangled states in a Hilbert space up to a dimension of sixteen by using all the four resonators, and do\textcolor{black}{es} not \textcolor{black}{require the} use of auxiliary rings for pump suppression \cite{liscidini2019scalable,clementi2023programmable}. 

We first focus on the generation of two-dimensional Bell states, and demonstrate that the source brightness is independent of the bin spacing. 
Then we show that the chip can be programmed to generate different qutrit and ququart target states, \textcolor{black}{some of which cannot} 
be directly generated by a single resonator, and we validate their fidelity using quantum state tomography. From the reconstructed density matrices, we certify the qudit dimension through device independent dimension witness \cite{wang2018multidimensional}, and the presence of entanglement from the violation of the generalized Collins-Gisin-Linded-Massar-Popescu (CGLMP) inequalities \cite{collins2002bell}. Last, with the perspective of implementing our device  for secure communication, we measure the correlation matrices of maximally entangled two\textcolor{black}{-} and three\textcolor{black}{-}dimensional Bell states on a set of mutually unbiased \textcolor{black}{bases.} 
We show fidelities exceeding $98\%$, comparable to those obtained for path-encoded qudits \cite{ding2017high,wang2018multidimensional}, which certifies the high quality of the generated states and of their coherent control.
\section{Results}


\begin{figure*}
    \includegraphics[width=1\textwidth]{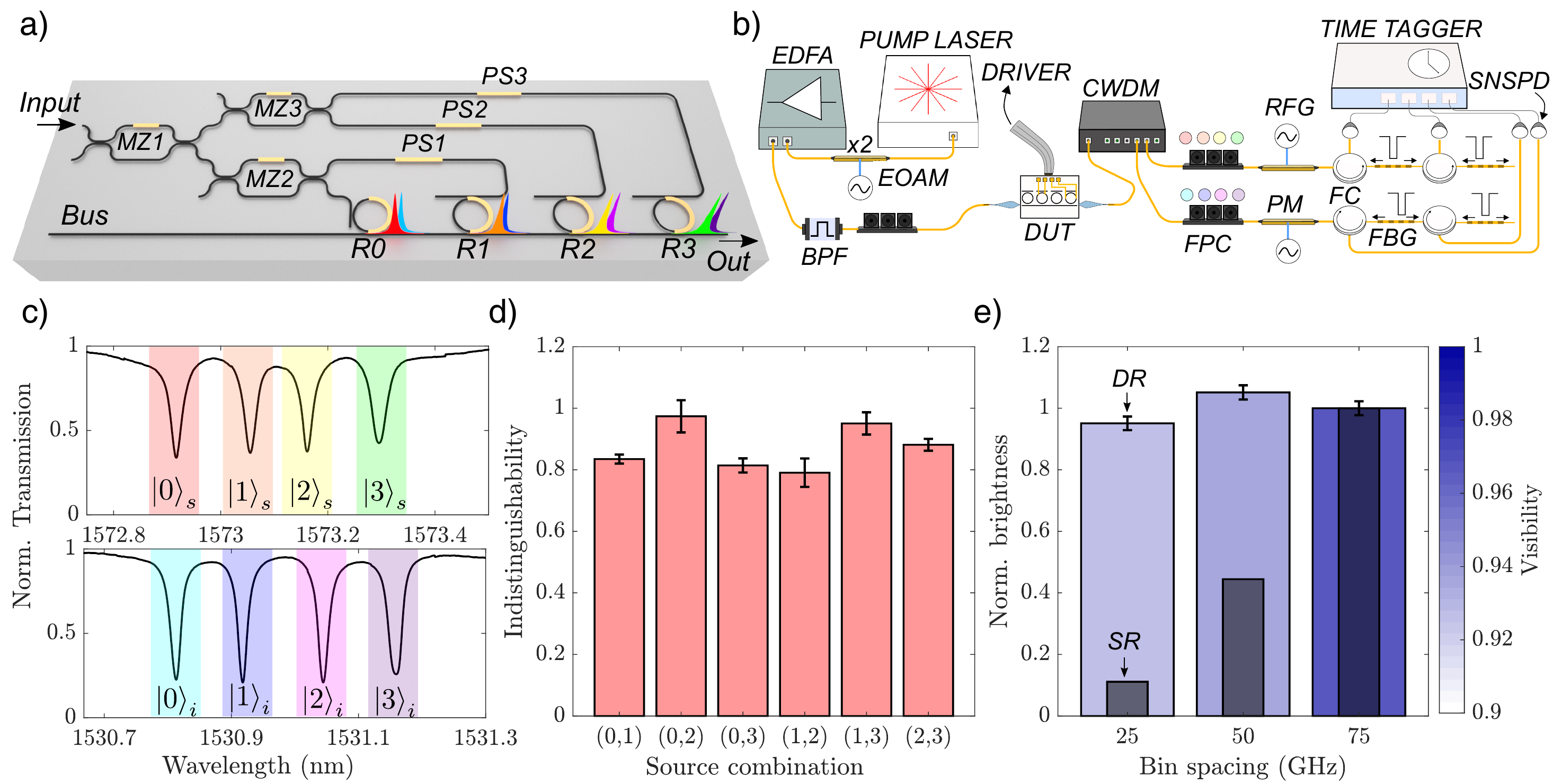}
    \caption{\label{fig:1} (a) Sketch of the circuit layout. Light is coupled to the chip from the Input port, and is split by a tree of Mach-Zehnder interferometers (MZ1-MZ3) into four paths, each exciting a microring resonator (R0-R3). The relative phases of the paths can be adjusted using the phase shifters PS1-PS3. Photon pairs are generated by SFWM over different frequency bins (colored pulses indicate a signal-idler pair). 
    (b) Sketch of the experimental setup. EDFA = Erbium-doped fiber amplifier, BPF = bandpass filter, EOAM = electro-optic amplitude modulator, DUT = device under test, CWDM = coarse wavelength division multiplexer, FPC = fiber polarization controller, RFG = radio frequency generator, PM = phase modulator, FC =  fiber circulator, FBG = fiber-Bragg-grating,  SNSPD = superconducting nanowire single photon detector. (c) Transmission spectra at the output waveguide. Signal and idler photons generated in different frequency bins are labeled as $\ket{k}_{s(i)}$, where $k=[0-3]$. A signal (idler) photon in bin $k$ can only be generated by the k$^{\textrm{th}}$ ring. (d) Indistinguishability between all the possible combinations of sources. 
    (e) Measured brightness of the double (DR) ring source, and simulated brightness of the single (SR) ring source, obtained by varying the frequency bin separation. Values are normalized to that at $75$ GHz. The visibility of the two photon interference is mapped into the bar color. }
\end{figure*}

Our source consists \textcolor{black}{of} 
a photonic circuit realized on the silicon-on-insulator (SOI) platform\textcolor{black}{; see Supplemental Material \cite{supp} for details of the fabrication procedure.} 
The chip is mounted on a sample holder that is temperature stabilized\textcolor{black}{, and metallic }
microheaters allow \textcolor{black}{us} to independently tune different parts of the circuit through the thermo-optic effect. 
Lensed fibers are used to couple \textcolor{black}{light into and out of} 
the chip through inverse tapers with a coupling loss of approximately $3.5$ dB/facet. 
A sketch of the layout of the device  is shown in Fig.\ref{fig:1}(a). A tree of Mach-Zehnder interferometers (MZ$1$-MZ$3$) splits light into four paths, with adjustable amplitude\textcolor{black}{s}, and  excites four identical add-drop microring resonators (R$0$-R$3$) of radius $22\,\mu$m ($\textrm{FSR}\simeq524$ GHz). The comb lines of each pair of rings $(i,j)$ are offset by an amount $\delta_{ij}=\delta|i-j|$, where $\delta$ is the tunable bin spacing. 
The transmission spectra at the output waveguide of the device is shown in Fig.\ref{fig:1}(c). It displays the four signal and idler resonances 
that encode the logical states $\ket{k}_{s(i)}$, \textcolor{black}{where $k=0,1,2,3$; here the bin spacing is set to $\delta=15$ GHz.} The average (loaded) quality factor\textcolor{black}{s} of the signal and of the idler resonances 
\textcolor{black}{are} respectively $Q_s\simeq5.7\times10^4$ and $Q_i\simeq7.8\times10^4$. 

We use four mutually coherent pumps, each resonant with a different ring, to trigger the generation of photon pairs.
The pumps are obtained by passing a continuous wave laser ($\lambda_p=1551.9$ nm) through a cascade of two electro-optic amplitude modulators (EOAMS), 
both driven by a 
sinusoidal radio-frequency (RF) signal at frequency $\delta$ (see Fig.\ref{fig:1}(b)). This process creates several sidebands that are equally spaced in frequency. We adjust the relative phase of the two RF signals\textcolor{black}{,} and the bias point of the two modulators\textcolor{black}{,} to roughly equalize the amplitude of the baseband to that of the first two pairs of sidebands. Four out of the five comb lines are used to pump the rings. 
Photon pair generation by SFWM occurs in all the rings in a coherent superposition. We operate at a pump power level such that multiple-pair events are negligible, which allows to write the output state as the two-photon state
\begin{equation}
    \ket{\Phi}=\frac{1}{\sqrt{D}}\sum_{k=0}^{D-1} \alpha_k \ket{k}_s\ket{k}_i, \label{eq:1}
\end{equation}
where $D$ is the number of frequency bins in the superposition and $\alpha_k$ are complex coefficients 
\textcolor{black}{that} depend on the pump properties, the MZi and the phase shifters PSi, as shown in Fig. 1(a). Fig.\ref{fig:1}(a). Photon pairs are pre-filtered from the pump using a coarse wavelength division multiplexer (CWDM), which also separates the signal from the idler. After that, they pass through 
separate EO phase modulators (PM), which are used to analyze the state. The PMs are driven by RF signals at a frequency $\delta$, which are both coherent with the one driving the EO of the pump. The power $P_{s(i)}$ and the phase $\theta_{s(i)}$ of the signal (idler) PM is varied in the state analysis.  
Tunable wavelength fiber-Bragg-gratings (FBG), with a bandwidth of $10$ GHz, are used to filter out photons from specific combinations of frequency bins, which are then detected by an array of superconducting nanowire single-photon detectors (SNSPDs). Coincidence measurements between the arrival time of the signal and the idler photon are performed using a time-tagging electronics. We used four FBGs and four detectors to simultaneously measure four different frequency bin combinations for each filter setting, which reduces the characterization time significantly.  


\subsection{Sources indistinguishability and flex-grid operation}
The two-photon state in Eq.(\ref{eq:1}) consists in a coherent superposition of $D$ frequency bins, each associated 
\textcolor{black}{with} a different source. \textcolor{black}{And each} 
has a different joint spectral amplitude $\Phi_i(\omega_s,\omega_i)$, which describes the intra-bin frequency correlation between signal and 
 idler photons. However, those fine-grain correlations are neglected in 
\textcolor{black}{the} simpler description (\ref{eq:1}) of the state, in which the frequency distribution is collapsed to a single label $\ket{k}_s\ket{k}_i$. It can be demonstrated that this simplified model reproduces the measurement outcomes if the elements $\rho_{jj,kk}$ of the density matrix $\rho=\ket{\Phi}\bra{\Phi}$ are replaced by $\rho_{jj,kk}\rightarrow I_{jk}\rho_{jj,kk}$, where $I_{jk}$ is the indistinguishability between source $i$ and $j$ (see Supplemental Material \cite{supp} for details). For $I_{jk}<1$, the modified density matrix describes a mixed state, as the fine-grain correlations depends on the specific bin. Thus the 
$I_{jk}$ are key parameters to characterize the state. 
\textcolor{black}{The fine-grain indistinguishability} can be measured by equally pumping two rings, \textcolor{black}{say $j$ and $k$, to generate the maximally entangled two-qubit state \mbox{$\ket{\psi}=\big(\ket{j}_s\ket{j}_i+\ket{k}_s\ket{k}_i\big)/{\sqrt{2}}$}, and by recording a two-photon interference fringe from the projection of the signal and the idler photon on $\bra{j}_{s(i)}+e^{j\theta_{s(i)}}\bra{k}_{s(i)}$; 
$I_{jk}$  
can be extracted from the visibility $V_{jk}$ of the fringe,} \textcolor{black}{where for a qubit 
$V_{jk}=I_{jk}$ \cite{kues2017chip}.}

For this experiment, the bin spacing 
\textcolor{black}{are} set to $15$ GHz. The signal/idler projectors 
\textcolor{black}{are} realized by adjusting the RF power driving the PMs to equalize the intensity $|J_{\pm 1}|$ of the first-order sidebands to that of the baseband ($|J_0|$)\textcolor{black}{; here the $J_n$ are Bessel functions of the first kind, with their argument equal to the strength of the modulation index (see Supplemental Material \cite{supp}).} The RF phase $\theta_s=\theta_i=\bar{\theta}$ 
\textcolor{black}{is} varied to record a two-photon interference coincidence measurement, in which the count rate is proportional to $\propto 1+V\textrm{cos}(2\bar{\theta})$  \cite{kues2017chip, imany201850}. 
The measured values of $I_{jk}$ are reported in Fig.\ref{fig:1}(d). The indistinguishability values are within the interval $[0.8-1]$, with an average value of  $V=0.87(7)$. This means that the joint spectral amplitudes of the different sources do not perfectly overlap. This is not surprising, for the resonances of the four resonators slightly differ in both their quality factors and extinctions, most likely due to small inhomogeneities in the waveguide-resonator coupling gap (almost at the very limit of the lithographic tool).


\textcolor{black}{We now turn to} 
the brightness of each resonator, which has an average value of \textcolor{black} {$B=(0.63\pm 0.15) \textrm{MHz} / (\textrm{mW})^{2}$}. Unlike the case of a single ring \cite{liscidini2019scalable}, in our system the brightness does not depend on the choice of the bin spacing. For example, we consider the Bell state $\ket{\psi}$ generated by pumping resonators R2 and R3 and 
measure the sum of the brightness of the two sources for three different bin spacing $\delta$: $25$ GHz, $50$ GHz and $75$ GHz. The results, shown in Fig.\ref{fig:1}(e), are normalized to the brightness level measured at a bin separation of $75$ GHz. For each configuration, we also calculated the brightness of a single ring with FSR=$\delta$ and with the same Q-factor. The data for a single ring are normalized to the brightness level calculated at a bin separation of $75$ GHz. For the double ring structure, the variation of the brightness between the different configurations is negligible. Instead, if a single ring were used, there would be a reduction of almost one order of magnitude 
\textcolor{black}{between the smallest and the largest bin separation \textcolor{black}{($(25/75)^{2}\simeq 0.11$)},} due to the intrinsic trade-off between brightness and ring finesse \cite{liscidini2019scalable}. Note that 
a bin spacing of $15$ GHz in a single ring with the same Q-factor would correspond to a generation rate \textcolor{black}{$(15/524)^{2} \sim 10^{-3}$} 
times smaller than that of a double ring source. Moreover, the single ring would have a radius \textcolor{black}{$R=({v_g}/(2\pi\delta)) \sim 800\, \mu$m,}  
where $v_g$ is the group velocity, occupying a footprint 
$\sim10^3$ times larger than that of the four rings.\par
The bin spacing 
\textcolor{black}{can} be reconfigured (flex-grid operation) without compromising the quality of entanglement. We prove this by extracting the visibility of quantum interference for each bin spacing\textcolor{black}{; under} 
the assumption of the presence of a color-less noise in our system, \textcolor{black}{this} can be used to certify  entanglement \cite{collins2002bell}. Following the 
approach of \textcolor{black}{Imany et al. }\cite{imany201850}, we set the driving frequency of the PMs to be half of the bin spacing and 
collect coincidences in the bins 
half-way between $\ket{2}_{s(i)}$ and $\ket{3}_{s(i)}$. 
The fringe visibilities $V$ are reported in Fig.\ref{fig:1}(e). These are all above $95\%$, and they do not depend on the bin separation. Moreover, being greater than $1/\sqrt{2}$, 
they also indicate entanglement in the generated state  \cite{kues2017chip}.   

\subsection{Programmable generation of qudit states}
We now demonstrate the programmability of our device by generating a variety of entangled qudit states that we characterize through quantum state tomography. In what follows, the bin spacing is fixed to $15$ GHz. Quantum state tomography is performed with $N$ measurement settings $\bm{\mathcal{M}(\bm{\gamma})}=\{\mathcal{M}_i(\bm{\gamma_i})\}_{i=i,..,N}$, each associated 
\textcolor{black}{with} a set of $D^2$ positive operator-valued measures (POVM) $\{\mathcal{P}_{iq}\}_{q=1,..,D^2}$ that describe all the possible experimental measurement outcomes, \textcolor{black}{where the probability of obtaining the $q^{\textup{th}}$ outcome for the $i^{\textup{th}}$ measurement setting is} 
$\textrm{Tr}(\rho\mathcal{P}_{iq})$. We use $\bm{\gamma}_i$ to indicate the $i^{\textrm{th}}$ measurement setting (RF power and phase) of the signal and the idler PM. The POVMs $\mathcal{P}_{iq}$ are defined as \mbox{$\mathcal{P}_{iq}=U^{\dagger}(\bm{\gamma}_i)\ket{q}\bra{q}U(\bm{\gamma}_i)$}, where $U(\bm{\gamma}_i)$ describes the action of the PMs on the system and $\ket{q}$ indicates one of the $D^2$ computational basis vectors. We used $N=5$ measurement settings for $D=2$, and $N=17$ for $D=3$ and $D=4$, corresponding respectively to $20\,(D=2)$, $153\,(D=3)$ and $272\,(D=4)$ POVMs. These are more than the $D^4$ measurements required to unambiguously determine the density matrix, but the redundancy helps to speed up the convergence of the reconstruction algorithm (maximum likelihood). The complete list of the $\bm{\gamma}_i$ is reported in the \textcolor{black}{Supplemental Material} 
\cite{supp}.
%

\begin{figure*}[t!]
  \centering
  \includegraphics[scale = 0.75]{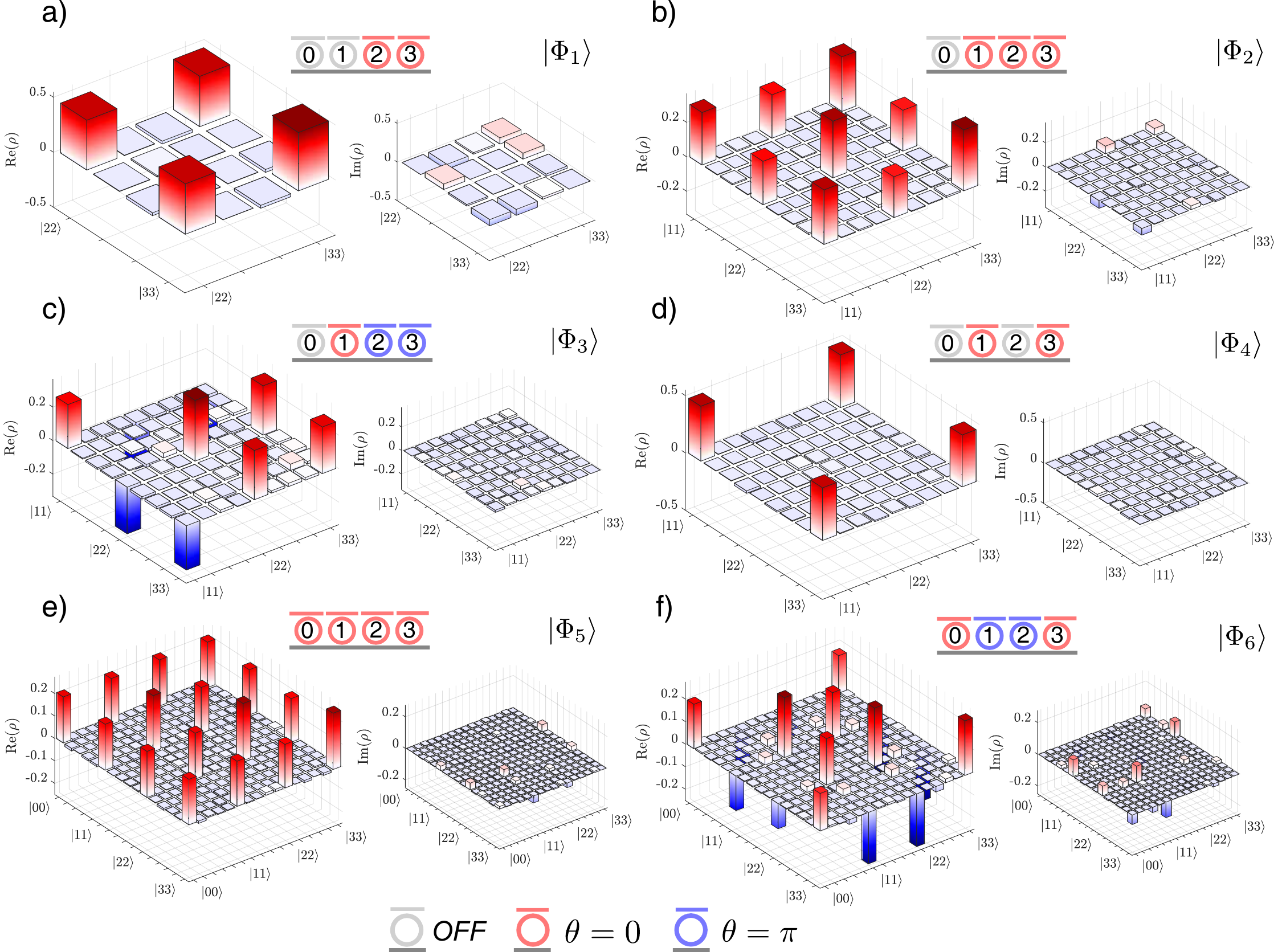}
\caption{(a-f): Reconstructed density matrices (the real and the imaginary parts are reported respectively to the left and to the right of each panel) for the different target quantum states reported in Table \ref{Table_1}. 
The sketch on the top of each matrix indicates the status of the four rings R0-R3 (the gray color corresponds to an off-resonance source) and the phase of each term in the quantum states in Table \ref{Table_1}.  \label{Figure_2}}
\end{figure*}

\textcolor{black}{We first program the two-qubit Bell state \textcolor{black}{$\ket{\Phi_1}=(\ket{22}+\ket{33})/\sqrt{2}$} by coherently pumping rings R2 and R3}.
The reconstructed density matrix $\rho_r$ is shown in Fig.\ref{Figure_2}(a). 
The state purity is $\mathcal{P}_1 = 94.63(4)\%$ and the fidelity $\mathcal{F}_1=\textrm{Tr}\sqrt{\sqrt{\rho_{t}}\rho_{r}\sqrt{\rho_{t}}}$ with respect to the target density matrix $\rho_{t}=\ket{\Phi_1}\bra{\Phi_1}$ 
is $95.0(2)\%$.

\textcolor{black}{We then} 
program the device to generate the two qutrit states $\ket{\Phi_2}=(\ket{11}+\ket{22}+\ket{33})/\sqrt{3}$  
and $\ket{\Phi_3}=(\ket{11}-\ket{22}-\ket{33})/\sqrt{3}$.   
Note that $\ket{\Phi_3}$ cannot be directly generated using a single ring resonator without a subsequent state manipulation, 
for example by using a waveshaper. On the contrary, in our device, the state can be built ``piece by piece" by controlling the contributions of each generating ring. Indeed, \textcolor{black}{the states $\ket{\Phi_2}$ and $\ket{\Phi_3}$} 
are \textcolor{black}{obtained} 
by coherently pumping the rings R1, R2, and R3. \textcolor{black}{For} 
$\ket{\Phi_2}$, the relative phases between the biphoton emissions from all the sources are \textcolor{black}{set} equal, while \textcolor{black}{for} 
$\ket{\Phi_3}$ the emission from source $1$ \textcolor{black}{is set to have} 
a phase shift of $\pi$ with respect to those of sources $2$ and $3$. The phases are regulated through the on-chip phase shifters PS1 and PS2 in Fig.\ref{fig:1}(a)\textcolor{black}{. More} 
details on the calibration procedure are given in the \textcolor{black}{Supplemental Material} 
\cite{supp}\textcolor{black}{.}

The reconstructed density matrices are shown in Fig.\ref{Figure_2}(b-c). The fidelity with the target states $\ket{\Phi_2}$ and $\ket{\Phi_3}$ are respectively $\mathcal{F}_2=86.2(3)$ and $\mathcal{F}_3=86.1(3)$, while the purities are respectively $\mathcal{P}_2=78.3(5)$ and $\mathcal{P}_3=79.3(5)$, and are limited by the indistinguishability $I_{jk}$ between the sources. Indeed, if one considers an average value of $0.87(7)$ for $I_{jk}$, which is the result shown in Fig.\ref{fig:1}(d), our simulations indicate that the maximum attainable fidelity and purity for both states are $\mathcal{F}_{\textrm{max}}=91(5)\%$ 
and $\mathcal{P}_{\textrm{max}}=84(8)\%$ respectively; 
see \textcolor{black}{Supplemental Material} 
\cite{supp}. In addition, we believe that low purity values might be due to an imperfect calibration of the PMs and to drifts of the system during the data acquisition. 

The programmability of our source can be exploited to dynamically change the number of terms in the Bell state that is being generated. As an example, we first programmed the device to output the state $\ket{\Phi_2}$, and then we implemented the transformation $\ket{\Phi_2}\rightarrow \ket{\Phi_4}$, where $\ket{\Phi_4}=(\ket{11}+\ket{33})/\sqrt{2}$, 
by detuning R2 from the resonance condition. To reconstruct the density matrix of this state, we used the same measurement settings implemented for $\ket{\Phi_2}$, which allows \textcolor{black}{us} to evaluate the magnitude of residual component of the state $\ket{22}$ in $\ket{\Phi_4}$. The reconstructed density matrix is reported in Fig.\ref{Figure_2}(d), showing a fidelity of $\mathcal{F}_4=93.2(2)\%$ with the target state, and a purity of $\mathcal{P}_4=88.5(4)\%$. The elements $\rho_{22,jj}$ are comparable to the noise floor.

As a last example, we programmed the device to generate the two-ququart  
entangled states \textcolor{black}{$\ket{\Phi_5}=(\ket{00}+\ket{11}+\ket{22}+\ket{33})/\sqrt{4}$} 
and \textcolor{black}{$\ket{\Phi_6}=(\ket{00}-\ket{11}-\ket{22}+\ket{33})/\sqrt{4}$; 
note} that the state $\ket{\Phi_6}$ cannot be generated by a single resonator. The two-ququart states are obtained by  pumping all 
four rings R0-R3 simultaneously, and by regulating the relative phase of their biphoton emission using the phase shifters PS1, PS2, and PS3 in Fig.\ref{fig:1}(a). 
The reconstructed density matrices are reported in Fig.\ref{Figure_2}(e-f). We computed a fidelity of $\mathcal{F}_5=84.5(2)\%$ for $\ket{\Phi_5}$, and $\mathcal{F}_6=79.0(1)\%$ for $\ket{\Phi_6}$, with purities of respectively $\mathcal{P}_5=74.4(4)\%$ and $\mathcal{P}_6=76.2(1)\%$.   

\textcolor{black}{Finally, we} 
used the reconstructed density matrices to \textcolor{black}{confirm that} 
the generated states violate the generalized CGLMP inequality for qudits \cite{collins2002bell}, which \textcolor{black}{verifies} 
the presence of entanglement.
The calculated CGLMP parameter $S$ is reported in Table \ref{Table_1}. For all the states we find that $S>2$, hallmarking the presence of non-local correlations. \textcolor{black}{One can also investigate the}
local qudit dimension through device independent dimension witness. As explained 
by Wang et al. \cite{wang2018multidimensional}, it is possible to calculate a lower bound $\lceil \mathcal{D}\rceil$ (where $\lceil x \rceil$ indicates the least integer greater than $x$) 
\textcolor{black}{of} the dimensionality of each subsystem from only local measurements in the $Z$-basis (i.e., the computational basis). Following this approach, we computed the values of $\mathcal{D}$ shown in Table \ref{Table_1}, which certify a local dimension of $\lceil \mathcal{D}\rceil=2$ for $\ket{\Phi_{1,4}}$, $\lceil \mathcal{D}\rceil=3$ for $\ket{\Phi_{2,3}}$, and $\lceil \mathcal{D}\rceil=4$ for $\ket{\Phi_{5,6}}$. 
\begin{figure*}[t!]
  \centering
  \includegraphics[scale = 0.6]{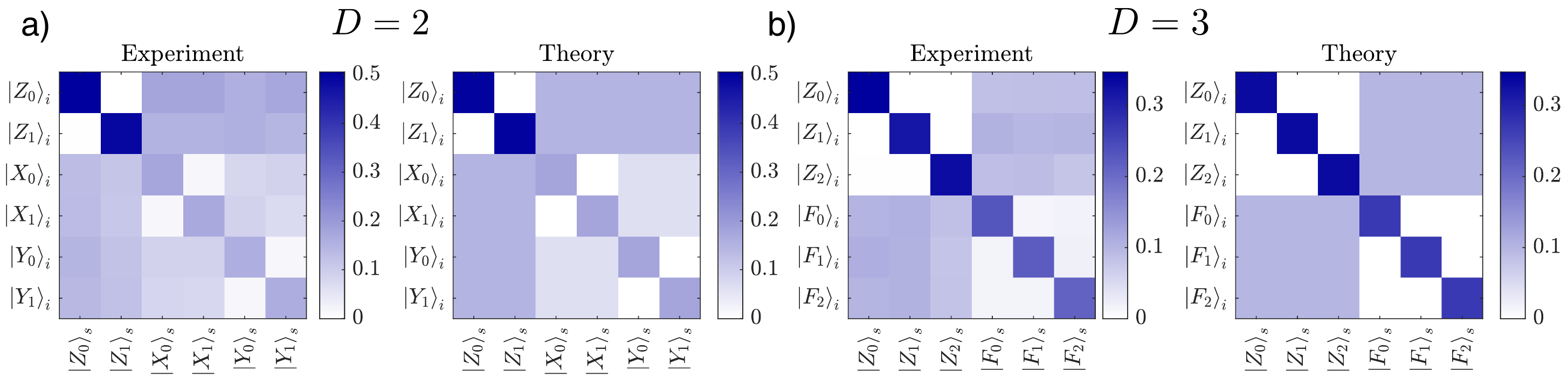}
\caption{(a) Experimental (left panel) and theoretical (right panel) distribution of the outcomes associated with the projections of the Bell state $\ket{\Phi_1}$ on a set of three mutually unbiased basis. These are the eigenvectors of the Pauli $X$, $Y$ and $Z$ operators. (b) Same as in panel (a), but relative to the three-dimensional Bell state $\ket{\Phi_2}$. The two sets of mutually unbiased basis are the eigenvectors of the Pauli $Z$ operator and the Fourier basis $F$. \label{Figure_3}}
\end{figure*}
\begin{table}[h!]
\begin{tabular}{cllllc}
Label & $\{\alpha_{0},\alpha_{1},\alpha_{2},\alpha_{3}\}$ & Fidelity & Purity & CGLMP & $\mathcal{D}$\\
\hline 
$\ket{\Phi_{1}}$ & $\{0,0,1,1\}$ & $95.0(2)$ & $94.63(4)$ & $2.59(1)$ & $1.882(5)$\\
$\ket{\Phi_{2}}$ & $\{0,1,1,1\}$ & $86.2(3)$ & $78.3(5)$ & $2.37(1)$ & $2.948(2)$\\
$\ket{\Phi_{3}}$ & $\{0,1,-1,-1\}$ & $86.1(3)$ & $79.3(5)$ & $2.337(7)$ & $2.705(5)$\\
$\ket{\Phi_{4}}$ & $\{0,1,0,1\}$ & $93.2(2)$ & $88.5(4)$ & 2.74(1) & $1.947(2)$\\
$\ket{\Phi_{5}}$ & $\{1,1,1,1\}$ & $84.5(2)$ & $74.4(4)$ & $2.336(7)$ & $3.747(5)$\\
$\ket{\Phi_{6}}$ & $\{1,-1,-1,1\}$ & $79.0(1)$ & $76.2(1)$ & $2.066(4)$ & $3.883(4)$\\
\end{tabular}
\caption{\label{Table_1}Values of the fidelity, purity, CGLMP parameter and dimension witness $\mathcal{D}$ for the target states \mbox{$\Phi_{j}$ ($j=[1-6]$)}. Each state is defined by the list of coefficients $\alpha_k$ as $\frac{1}{\sqrt{D}}\sum_{k=0}^{D-1}\alpha_k\ket{kk}$ . Errors are computed by assuming a Poissonian
statistics on the recorded coincidences. }
\end{table}

\subsection{Qubit and qutrit correlations on mutually unbiased basis}
Given two orthonormal basis sets $\{ \ket{e_1},...,\ket{e_D}\}$ and $\{ \ket{f_1},...,\ket{f_D}\}$, these are said to be mutually unbiased if $\braket{e_j|f_k}={D}^{-1}$. 
The generation of entangled states and the ability \textcolor{black}{of} 
two users to perform projections on a set of mutually unbiased basis are the backbone of many secure communication protocols \cite{cerf2002security}. 
Maximally entangled Bell states have the property that the outcomes of the measurements between two parties, say Alice and Bob, are always perfectly correlated or anti-correlated when they measure 
\textcolor{black}{by using} the same basis. This allows 
users to share the same secret key in a communication protocol. At the same time, the two outcomes are completely uncorrelated when Alice and Bob choose two 
 mutually unbiased basis sets. More formally, given a maximally entangled Bell state $\ket{\Phi}$, we have that $|\braket{e_j e_k^{*}|\Phi}|^2={D}^{-1}\delta_{jk}$ 
and $|\braket{e_jf_k^{*}|\Phi}|^2={D}^{-1}$,  
where the states $\ket{f^{*}},\ket{e^{*}}$ are obtained from the complex conjugation of the expansion coefficients of $\ket{f},\ket{e}$ on the computational basis.
An important parameter that allows \textcolor{black}{one} to predict an upper bound to the tolerable level of errors in the communication, and which defines the amount of secure bits that can be extracted from the secret key, is the fidelity $\mathcal{F}_{\textrm{mub}}$ with which such correlations are reproduced in the experiment \cite{cerf2002security,ding2017high}.

We evaluate this parameter for our entangled photon source by using the PMs to perform projective measurements on mutually unbiased basis. We choose to work with the qubit and qutrit states $\ket{\Phi_1}$ and $\ket{\Phi_2}$. For the qubit, we choose the eigenvectors of the Pauli operators $X$, $Y$, and $Z$  as a set of three mutually unbiased basis. The projections on the $X$ and $Y$ basis are implemented by first adjusting the RF power of the PMs to have $|J_0|=|J_{\pm 1}|=\bar{J}$, and then by choosing the RF phases within the set $\{0,\pi\}$ ($X$ set) or $\{0,\pi/{2}\}$ 
($Y$ set). We collect coincidences from the signal/idler bins $\ket{2}$. In the case of the qutrit, we only implement two out of the four mutually unbiased basis sets \cite{lu2020three}, which we choose to be the $Z$ and the Fourier basis \cite{wang2018multidimensional}. The remaining sets cannot be implemented with a single PM, since they require \textcolor{black}{photon projection on} 
bin superpositions 
\textcolor{black}{that} have a relative phase that is not monotonic with the bin number \cite{lu2020three}. However, they could be realized with a QFP \cite{lu2022high}.  
The Fourier basis is realized by setting the PMs to have $|J_0| = |J_{\pm1}|$, and then by varying the RF phases within the set
$\{0,2\pi/3,4\pi/3\}$.
Coincidences are collected in the signal/idler bin $\ket{1}$. Because $J_{-n}=(-1)^nJ_{n}$, we actually implement $I^{-}F$, where $I^{-}$ is a $3\times 3$ identity matrix except that  $I^{-}_{33}=-1$, and $F$ is the Fourier matrix \cite{lukens2017frequency}.
The measured correlation matrices $C_e$ for dimension $D=2$ (qubit) and $D=3$ (qutrit) are respectively shown in Fig.\ref{Figure_3}(a) and in Fig.\ref{Figure_3}(b). The 
\textcolor{black}{fidelity}  $\mathcal{F}_{\textrm{mub}}$ with the theoretical expectations $C_t$ is defined as \cite{lukens2019all}
\begin{equation}
\mathcal{F}_{\textrm{mub}}\equiv\frac{\textrm{Tr}({C_e}^{\dagger}C_t)\textrm{Tr}({C_t}^{\dagger}C_e)}{\textrm{Tr}(C_e^{\dagger}C_e)\textrm{Tr}(C_t^{\dagger}C_t)}.    
\end{equation}
We find $\mathcal{F}_{\textrm{mub}}=98.76(3)\%$ for $D=2$, and \mbox{$\mathcal{F}_{\textrm{mub}}=98.78(3)\%$} for $D=3$,  which are values comparable to those obtained for path-encoded qudits \cite{ding2017high,wang2018multidimensional}. 

As shown in Fig.\ref{Figure_3}, the probabilities of the outcomes associated with the superposition basis $X$, $Y$ and $F$ do not sum to one. This happens because the PMs also scatter photons outside the computational basis, so that the implemented projectors $\ket{e'}=\bar{J}\sqrt{D}\ket{e}$ (where $\ket{e}$ can be any of the eigenvectors of the operators $X$, $Y$ or $F$) are not normalized to one. This does not introduce any bias in the correlation matrices in Fig.\ref{Figure_3}, and it acts only as an excess loss. 

More generally, the transformation imparted by a PM is described by a circulant matrix $W$ defined on the infinite-dimensional space spanned  by equidistant frequency bins \cite{lukens2019all}. Consequently, if $W$ changes the measurement basis to $Q=\{ \ket{q_0},...,\ket{q_{D-1}}\}=\{ W^{\dagger}\ket{0},...,W^{\dagger}\ket{D-1}\}$, the coefficients $W_{ik}$ of the expansion $\ket{q_i}=\sum_{k}W_{ik}\ket{i}$ are related by $W_{ik}=W_{i+m,k+m}$ (where $m$ is an integer number). 
Within the three-dimensional space \emph{span}$\{\ket{0},\ket{1},\ket{2} \}$, the PM can implement the transformation $W^{\dagger}\ket{1}=J_{-1}\ket{0}+J_0\ket{1}+J_{1}\ket{2}=\bar{J}\sqrt{3}\ket{F_0}$ ($\ket{F_0}=\left ( \ket{0}+\ket{1}-\ket{2}\right )/\sqrt{3}$), which allows us to estimate the probability  $\bra{F'_0}\rho\ket{F'_0}$ from the photon counts measured on $\ket{1}$, but the same PM setting will also implement $W^{\dagger}\ket{0}=J_0\ket{0}+J_{1}\ket{1}+J_2\ket{2}$ and $W^{\dagger}\ket{2}=J_{-2}\ket{0}+J_{-1}\ket{1}+J_0\ket{2}$, which differ with respect to both $\ket{F_1}=\left (e^{-i\frac{2\pi}{3}}\ket{0}+\ket{1}-e^{i\frac{2\pi}{3}}\ket{2}\right )/\sqrt{3}$ and $\ket{F_2}=\left (e^{-i\frac{4\pi}{3}}\ket{0}+\ket{1}-e^{i\frac{4\pi}{3}}\ket{2}\right )/\sqrt{3}$ (e.g., $J_{-2}$ cannot be equal to $J_{-1}$ and $J_0$ simultaneously). As a consequence,
we can apply $W$ to map any of the (not normalized) vectors $\ket{F'_j}=\bar{J}\sqrt{D}\ket{F_j}$ to a particular $Z$-basis vector, in this case $\ket{1}$, but $W$ will not map $\ket{F'_{j\neq k}}$ to $\ket{0}$ and $\ket{2}$;  these basis vectors can not be  measured without changing the PM setting.
It will be then very inefficient to run a communication protocol using only PMs, but these limitations can be overcome through the use of QFPs, which can be used to implement arbitrary qubit gates \cite{lu2020fully} and the high-dimensional Fourier gate \cite{lu2022high} with almost unit probability. The process fidelity of those gates exceeds $99.9\%$, which suggests that high-dimensional entanglement-based QKD could be implemented by using our source.

\section{\label{sec:Conclusion} Conclusion}
We presented a programmable silicon photonics chip for the generation of frequency-bin entangled qudits up to a dimension of four. Each bin is associated 
\textcolor{black}{with} the spontaneous emission of photon pairs from different high-finesse ring-resonators, thus decoupling the generation rate from the spectral separation of the bins. The device can be programmed to generate multiple qudit entangled states, including some that cannot be directly generated using a single ring. All the generated states were validated by quantum state tomography. We measured the correlation matrices of maximally entangled two-qubit and two-qutrit states when 
projected on different mutually unbiased basis sets, finding a high fidelity with the target correlation matrices. This indicates that our source could be used in high dimensional secure communication protocols, and we \textcolor{black}{can} 
envision other applications, such as the multiplexing of single photons.

Indeed, a variable electro-optic frequency shift can be used as a feed-forward operation to increase the probability of having a single photon at the output without a corresponding increase of multiphoton events \cite{puigibert2017heralded}. A similar approach has been used to herald single photons in pure spectral modes \cite{hiemstra2020pure}. 
In both cases, the number of multiplexed spectral modes is limited by the maximum 
frequency shift that can be imparted to the photons. It is then important to have a source that generates closely spaced and bright spectral modes, which are exactly what our device can provide.

Furthermore, we could exploit the programmability of our source in  quantum simulations and multiphoton interference. For example, quantum transport of correlated biphotons in discrete frequency modes 
\textcolor{black}{exhibits} a rich variety of behaviors, from walks showing enhanced ballistic transport to subspaces featuring scattering centers or local traps \cite{imany2020probing}. Interestingly, one can change from a diffusion regime to another by modifying the relative phases between the comb lines, which in our case 
can be done on chip. 
Since the walk-depth increases with the strength of the electro-optic modulation, large depths could be reached by using comb lines that are closely spaced, for which the electro-optic mixing is more effective. 
\textcolor{black}{Very generally,} the high brightness of our source could be exploited to probe multiphoton interference in the frequency domain, in analogy with boson sampling experiments previously reported in path \cite{paesani2019generation} and time-bin encoding \cite{madsen2022quantum}, or in frustrated interference \cite{gu2019quantum,hochrainer2022quantum}.
\acknowledgements{
This work has been supported by Ministero dell’Istruzione,
dell’ Università e della Ricerca [MIUR grant Dipartimenti
di Eccellenza 2018-2022 (F11I18000680001)]. This work is partly supported by the French RENATECH network. J. E. S.
acknowledges support from the Natural Sciences and 
Engineering Research Council of Canada. The device has
been designed using the open source Nazca design\texttrademark \,framework.}
\newpage
\nocite{bellegarde2018improvement,birge2003psot,lu2022bayesian,seshadri2022nonlocal,helt2010spontaneous}
%

\end{document}


\preprint{APS/123-QED}

\title{Supplemental material for A reconfigurable silicon photonics chip for the generation of frequency bin entangled qudits}

\author{Massimo Borghi}
\email{massimo.borghi@unipv.it}
\affiliation{Dipartimento di Fisica, Università di Pavia, Via Agostino Bassi 6, Pavia 27100, Italy}

\author{Noemi Tagliavacche}
\affiliation{Dipartimento di Fisica, Università di Pavia, Via Agostino Bassi 6, Pavia 27100, Italy}

\author{Federico Andrea Sabattoli}
\altaffiliation[Current address:]{ Advanced Fiber Resources Milan S.r.L, via Fellini 4, 20097 San Donato Milanese (MI), Italy.}
\affiliation{Dipartimento di Fisica, Università di Pavia, Via Agostino Bassi 6, Pavia 27100, Italy}

\author{Houssein El Dirani}
\altaffiliation[Current address:]{ LIGENTEC  SA,  224  Bd  John  Kennedy, 91100 Corbeil-Essonnes, France.}
\affiliation{Université Grenoble Alpes, CEA-Leti, Grenoble cedex 38054, France}

\author{Laurene Youssef}
\altaffiliation[Current address:]{ IRCER, Centre Européen de la Céramique, 12 rue Atlantis, 87068 Limoges, France.}
\affiliation{Univ. Grenoble Alpes, CNRS, CEA/LETI-Minatec, Grenoble INP, LTM, F-38054 Grenoble-France}

\author{Camille Petit-Etienne}
\affiliation{Univ. Grenoble Alpes, CNRS, CEA/LETI-Minatec, Grenoble INP, LTM, F-38054 Grenoble-France}

\author{Erwine Pargon}
\affiliation{Univ. Grenoble Alpes, CNRS, CEA/LETI-Minatec, Grenoble INP, LTM, F-38054 Grenoble-France}

\author{J.E. Sipe}
\affiliation{Department of Physics, University of Toronto, 60 St. George Street, Toronto, Ontario M5S 1A7, Canada}


\author{Marco Liscidini}
\affiliation{Dipartimento di Fisica, Università di Pavia, Via Agostino Bassi 6, Pavia 27100, Italy}

\author{Corrado Sciancalepore}
\altaffiliation[Current address:]{ SOITEC   SA,   Parc   Technologique   des Fontaines, Chemin des Franques, 38190 Bernin, France.}
\affiliation{Université Grenoble Alpes, CEA-Leti, Grenoble cedex 38054, France}

\author{Matteo Galli}
\email{matteo.galli@unipv.it}
\affiliation{Dipartimento di Fisica, Università di Pavia, Via Agostino Bassi 6, Pavia 27100, Italy}

\author{Daniele Bajoni}
\affiliation{Dipartimento di Ingegneria Industriale e dell’Informazione, Università di Pavia, Via Adolfo Ferrata 5, Pavia 27100, Italy 
}%

\date{\today}

\maketitle
\section{Experimental methods}
\label{Section: dev_fab}
\subsection{Sample fabrication}
\noindent The device was fabricated at CEA-LETI (Grenoble) on a 200 mm silicon-on-insulator wafer manufactured by SOITEC (Bernin) with a 220 nm-thick device layer of crystalline silicon. Silicon structures are patterned by deep ultraviolet lithography with 120 nm resolution, followed by inductively coupled plasma etching (HBr/Cl$_2$/He-O$_2$, realized in collaboration with LTM (Laboratoire
des Technologies de la Microelectronique)) and post-etch $\textrm{O}_2$ plasma for resist stripping. The sidewall roughness of the waveguides is reduced by 
\textcolor{black}{hydrogen} annealing  \cite{bellegarde2018improvement}. After encapsulation resulting in a 900 nm-thick $\textrm{SiO}_2$ layer, 110 nm of titanium
nitride (TiN) 
\textcolor{black}{was} deposited and patterned to create the thermal phase
shifters. Aluminum copper (AlCu) was employed for the electrical pads.
A deep etch comprising two different steps ($\textrm{C}_4\textrm{F}_8/\textrm{O}2/\textrm{CO}/\textrm{Ar}$ plasma for the
whole depth of $\textrm{SiO}_2$, Bosch method to remove 150 $\mu$m of the 725 $\mu$m-thick Si
substrate) was implemented to separate the sub-dice and to realize high quality end facets for edge coupling.
\subsection{Device geometry}
\noindent The silicon waveguides have a cross section of $600\,\textrm{nm} \times 220\, \textrm{nm}$. 
\textcolor{black}{The ring} resonators have a radius of $22\,\mu$m\textcolor{black}{,} and a coupling gap with the bus waveguide of $150$ nm. The average separation between each resonator is $480\,\mu$m. The three Mach-Zehnder interferometers are each made by two $50/50$ splitters, which are realized using evanescent couplers (length $17.5\,\mu$m, coupling gap $120$ nm). The two arms have an equal length of $150\,\mu$m, which coincides with that of the heater. The phase shifters PS1, PS2 and PS3 in Fig.1(a) of the main text are realized by heaters of length $\sim 120\,\mu$m, which have a tuning efficiency of $\sim 15  \textrm{deg}/\textrm{mW}$. 

\subsection{On chip state reconfiguration}
\noindent Here we describe the step-by-step procedure that we implemented to generate the qutrit states $\ket{\Phi_2}=(\ket{11}+\ket{22}+\ket{33})/\sqrt{3}$ and $\ket{\Phi_3}=(\ket{11}-\ket{22}-\ket{33})/\sqrt{3}$, and the two-ququart states $\ket{\Phi_5}=(\ket{00}+\ket{11}+\ket{22}+\ket{33})/2$ and $\ket{\Phi_6}=(\ket{00}-\ket{11}-\ket{22}+\ket{33})/2$; see the main text for a description of the bin labels.
\textcolor{black}{The procedure} 
relies on the acquisition of two-photon interference fringes between adjacent pairs of frequency bins. These are obtained by projecting the signal and the idler photon on\textcolor{black}{to} $\bra{i}_{s(i)}+e^{i\theta_{s(i)}}\bra{i+1}_{s(i)}$, where $i$ labels one of the frequency bins. Those projectors are realized by adjusting the RF power driving the phase modulators (PM) in order to equalize the intensities $|J_{\pm 1}|$ of the first order sidebands to that of the baseband ($|J_0|$); \textcolor{black}{see below.} The RF phase $\theta_s=\theta_i=\bar{\theta}$ is varied to scan a two-photon interference fringe, which is proportional to $\propto 1+V\textrm{cos}(2\bar{\theta})$ ($V$ is the fringe visibility).\par
\emph{Qutrit states}: for the state  $\ket{\Phi_2}$, the relative phase between the biphoton emission from ring R1, R2, and R3 must be set equal. This is accomplished in two steps. First, we acquired a two-photon interference fringe by  pumping only R1 and R2 (interference fringe 1-2). Second, a fringe is acquired by pumping only R2 and R3 (interference fringe 2-3). 
We then exploited the phase shifter PS1 (see Fig.1(a) in the main text) to align the two interference fringes. 
Note that one could have chosen the maximum of the fringe 1-2 as a reference, and used PS2 (or both PS1 and PS2) to align the maximum of two 
 fringes. 
In the state $\ket{\Phi_3}$, the biphoton emission from R1 has a phase shift of $\pi$ with respect to those from R2 and R3. In this case, we aligned the minimum of the fringe 1-2 to the maximum of the fringe 2-3. 

\emph{Ququart states}: for these states we acquired three interference curves, one for each pair of adjacent rings (R0-R1, R1-R2, and R2-R3). 
To generate the state $\ket{\Phi_5}$, the relative phase between the biphoton emission from all the four rings must be set  equal. We then aligned the maxima of all the three interference curves through the phase shifters PS1, PS2,  and PS3. To generate the state $\ket{\Phi_6}$, we must set a relative phase of $\pi$ between the biphoton emission from R1 and R2 to those of R0 and R3. Using PS3 and PS2, we aligned the minimum of the fringe 2-3 to the maximum of the fringe 1-2. After that, using PS1 we aligned the minimum of the fringe 0-1 to the minimum of the fringe 2-3.
In both the qutrit and the ququart cases, this procedure is repeated two times to mitigate the effect of thermal cross-talk between the heaters.

\section{Density matrix reconstruction}
 
We implemented quantum state tomography to reconstruct the density matrix $\rho$ of the state by performing a series of projective measurements. The probability $p_q^{(j)}$ of getting one of the $D^2$ ($D$ is the dimension of each of the two qudits) measurement outcomes is given by

\begin{equation}
    p^{(j)}_q=\textrm{Tr}(\rho\mathcal{P}_{jq}),
    \label{equ:p_q}
\end{equation}

\noindent where $\mathcal{P}_{jq}$ is a positive-operator-valued-measure (POVM), defined as \mbox{$\mathcal{P}_{jq}=U^{\dagger}(\bm{\gamma}_j)\ket{q}\bra{q}U(\bm{\gamma}_j)$}. The operator $U(\bm{\gamma}_j)=U_s(P_s,\theta_s)\otimes U_i(P_i,\theta_i)$ describes the action of the electro-optic phase modulators (PMs) on the signal/idler photons, and $\ket{q}$ indicates one of the $D^2$ computational basis vectors.
We use $\gamma_j$ to label the $j^{\textrm{th}}$
measurement setting, which comprises the RF power $P_{s(i)}^{(\gamma_i)}$ and and phase $\theta_{s(i)}^{(\gamma_i)}$ of the signal and the idler PM.
Equation \ref{equ:p_q} can be rewritten as

\begin{equation}
    p_{mn}=\textrm{Tr}(\rho U^{\dagger}\ket{mn}\bra{mn}U)=\bra{mn}U\rho U^{\dagger}\ket{mn},
    \label{eq:p_mn}
\end{equation}

\noindent where for simplicity $U=U(\gamma_j)$. 
For each measurement setting $\gamma_j$=($P_s,\theta_s,P_i,\theta_i$), we collected coincidence counts $C_{mn}$ for all the possible $\{(m,n)\,|\,m\in [0,..,D-1], n\in [0,D-1]\}$ ($D$ is the qudit dimension) frequency bins combinations. The measurement settings used to reconstruct the density matrices are reported in Table \ref{Table_1}.  
The coincidence counts $C_{mn}$ recorded in the experiment are related to the probabilities $p_{mn}$ as

\begin{equation}
    C_{mn}= R\cdot L_{mn}\cdot p_{mn}, \label{eq:S1}
\end{equation}
 
\noindent where $R$ is the rate of photon pairs at the modulators and $L_{mn}$ are (measured) the losses from the modulators to the detectors. 
We used the particle swarm optimization algorithm \cite{birge2003psot} to minimize the following cost function $f_{\textrm{cost}}$
\begin{equation}
    f_{\textrm{cost}}=\frac{1}{n_{\textrm{meas}}}\sum_{k=0}^{n_{\textrm{meas}}-1} \frac{(C_{\textrm{theo}}[k]-C_{\textrm{exp}}[k])^{2}}{2 C_{\textrm{theo}}[k], }
    \label{fcost}
\end{equation}
where $n_{\textrm{meas}}$ is the number of measurements, $C_{\textrm{exp}}$ is the vector containing the recorded coincidence measurements and $C_{\textrm{theo}}$  is the vector of expected coincidence counts, which are calculated from a candidate density matrix $\rho$ by using Eq.(\ref{eq:S1}). The matrix $\rho$ is parameterized by a vector of $D^2$ real coefficients $\bm{t}$ as
\begin{equation}
    \rho(\bm{t}) = \frac{T(\bm{t})T^{\dagger}(\bm{t})}{\textrm{Tr}(T(\bm{t})T^{\dagger}(\bm{t}))},
\end{equation}
where $T$ is a complex lower triangular matrix with real elements on the diagonal. The particle swarm algorithm minimizes $f_{\textrm{cost}}$ in the parameter space spanned by $(R,\bm{t})$.
We used $N=5$ measurement settings for $D=2$, and $N=17$ for $D=3$ and $D=4$, corresponding respectively to $n_{\textrm{meas}}=20\,(D=2)$, $n_{\textrm{meas}}=153\,(D=3)$ and $n_{\textrm{meas}}=272\,(D=4)$. These are more than the $D^4$ measurements required to unambiguously determine the density matrix, but the redundancy helped to speed up the convergence of the reconstruction algorithm.

\begin{table}[h]
    \begin{subtable}[h]{0.6\textwidth}
        \begin{tabular}{l | l | l | l | l }
        \rowcolor[HTML]{FFCCC9} 
        \textbf{Measurement setting ($\gamma_j$)} & $\textbf{P}_\textbf{s}$ \textbf{(dBm)} & \textbf{$\theta_s$} \textbf{(rad)} & $\textbf{P}_\textbf{i}$ \textbf{(dBm)} & \textbf{$\theta_i$} \textbf{(rad)} \\
        \hline \hline
        $\gamma_1$ & $0$ & $0$ & $0$ & $0$ \\
        
        $\gamma_2$ & $22.1$ & $0$ & 24.3 & 0\\
        
        $\gamma_3$ & $22.1$ & $\frac{\pi}{2}$ & $24.3$ & $-\frac{\pi}{2}$\\
        
        $\gamma_4$ & 22.1 & $0$ & 0 & 0\\
        
        $\gamma_5$ & $0$ & $0$ & $24.3$ & $0$
       \end{tabular}
       \caption{}
       \label{tab_QB}
    \end{subtable}
    \renewcommand*{\arraystretch}{1.5}
    \begin{subtable}[h]{0.6\textwidth}
        \renewcommand*{\arraystretch}{1.5}
        \begin{tabular}{l | l | l | l | l }
        \rowcolor[HTML]{CBCEFB} 
        \textbf{Measurement setting ($\gamma_j$) } & $\textbf{P}_\textbf{s}$ \textbf{(dBm)} & \textbf{$\theta_s$} \textbf{(rad)} & $\textbf{P}_\textbf{i}$ \textbf{(dBm)} & \textbf{$\theta_i$} \textbf{(rad)} \\
        \hline \hline 
        $\gamma_1$ & $22.1$ & $0$ & $24.3$ & $0$ \\
        
        $\gamma_2$ & $22.1$ & $0$ & $24.3$ & $\frac{4\pi}{3}$ \\
        
        $\gamma_3$ & $22.1$ & $0$ & $24.3$ & -$\frac{4\pi}{3}$ \\
        
        $\gamma_4$ & $22.1$ & $0$ & $24.3$ & $\frac{2\pi}{3}$ \\

       $\gamma_5$ & $22.1$ & $\frac{4\pi}{3}$ & $24.3$ & 0 \\
        
        $\gamma_6$ & $22.1$ & $\frac{4\pi}{3}$ & $24.3$ & $\frac{4\pi}{3}$ \\
        
        $\gamma_7$ & $22.1$ & $\frac{4\pi}{3}$ & $24.3$ & -$\frac{4\pi}{3}$ \\
        
        $\gamma_8$ & $22.1$ & $\frac{4\pi}{3}$ & $24.3$ & $\frac{2\pi}{3}$ \\
        
        $\gamma_9$ & $22.1$ & -$\frac{4\pi}{3}$ & $24.3$ & $0$ \\
        
        $\gamma_{10}$ & $22.1$ & -$\frac{4\pi}{3}$ & $24.3$ & $\frac{4\pi}{3}$ \\
        
        $\gamma_{11}$ & $22.1$ & -$\frac{4\pi}{3}$ & $24.3$ & -$\frac{4\pi}{3}$ \\
        
        $\gamma_{12}$ & $22.1$ & -$\frac{4\pi}{3}$ & $24.3$ & $\frac{2\pi}{3}$ \\
        
        $\gamma_{13}$ & $22.1$ & $\frac{2\pi}{3}$ & $24.3$ & $0$ \\
        
        $\gamma_{14}$ & $22.1$ & $\frac{2\pi}{3}$ & $24.3$ & $\frac{4\pi}{3}$ \\
        
        $\gamma_{15}$ & $22.1$ & $\frac{2\pi}{3}$ & $24.3$ & -$\frac{4\pi}{3}$ \\
        
        $\gamma_{16}$ & $22.1$ & $\frac{2\pi}{3}$ & $24.3$ & $\frac{2\pi}{3}$ \\
        
        $\gamma_{17}$ & $0$ & $0$ & $0$ & $0$ 
        \end{tabular}
        \caption{}
        \label{tab_QT_QQ}
     \end{subtable}
     \caption{Signal and idler measurement settings $\gamma_j$ used to perform quantum state tomography on qubit (a) and qutrit-ququart (b) states. We report the RF powers $P_s$, $P_i$ and the phases $\theta_s$, $\theta_i$ of the signals driving the signal and the idler PM. The values $P_s=22.1$ dBm and $P_i=24.3$ dBm correspond to a modulation index of $\beta \sim 1.434$, for which the intensities of the first order sidebands $J_{\pm1}$ are equal to that of the baseband $J_0$ (the fact that $P_s<P_i$ is due to the higher modulation efficiency of the signal PM)  
     }
     \label{Table_1}
\end{table}

\subsection{Impact of source indistinguishability on the reconstruction of the density matrix} \label{sec.III}
The density matrix reconstruction is done through a series of projective measurements whose probability outcomes $p_{mn}$ on the bin combination $\{(m,n)\,|\,m\in [0,..,D-1], n\in [0,D-1]\}$ ($D$ is the qudit dimension) are given by Eq.(\ref{eq:p_mn}). The two-photon state $\ket{\Psi}$ generated by our device is:
\begin{equation}
    \ket{\Psi}=\sum_{k=0}^{D-1}\alpha_k \int \phi_{k}(\omega_{k_s}+\Omega_1,\omega_{k_i}+\Omega_2)\ket{\omega_{k_s}+\Omega_1,\omega_{k_i}+\Omega_2}d\Omega_1 d\Omega_2, \label{eq:state}
\end{equation}
where $\phi_{k}$ is the joint spectral amplitude of the $k^{\textrm{th}}$ ring resonator \cite{helt2010spontaneous}, which generates photon pairs in the signal/idler bins $\ket{k}_i\ket{k}_s$, and $\{\alpha\}_{k=0,..,D-1}$ is a list of complex coefficients such that $\sum_k |\alpha_k|^2=1$. The resonance frequency associated to the signal(idler) bin $k$ (see Fig.1(c) in the main text) is $\omega_{k_{s(i)}}$.
The density matrix $\rho$ then writes
\begin{equation}
    \rho = \sum_{\{k,k'\}=0}^{D-1}\int\alpha_k\alpha_{k'}^{*}\phi_{k}(\omega_{k_{s}}+\Omega_1,\omega_{k_{i}}+\Omega_2)\phi_{k'}^{*}(\omega_{k'_s}+\Omega_1',\omega_{k'_i}+\Omega_2')\ket{\omega_{k_s}+\Omega_1,\omega_{k_i}+\Omega_2}\bra{\omega_{k'_i}+\Omega_2',\omega_{k'_s}+\Omega_1'}
    d\Omega_1 d\Omega_1' d\Omega_2 d\Omega_2'. \label{eq:S5}
\end{equation}
Under the action of the PMs, the density matrix in Eq.(\ref{eq:S5}) evolves as $\rho \rightarrow U\rho U^{\dagger}$, where $U=U_i\otimes U_s$ transforms the signal/idler annihilation operator $a$ as
 \begin{equation}    U_{s(i)}a_m(\omega_{m_{s(i)}}+\omega)U_{s(i)}^{\dagger} = \sum_{r=-\infty}^{+\infty}V_{mr}^{s(i)}a(\omega_{r_{s(i)}}+\omega). \label{eq:S6}
 \end{equation}
The coefficients $V_{mr}^{s(i)}$ are related to the Bessel functions of first kind $J_{m-r}$, of order $r-m$, as
 \begin{equation}
 V_{mr}^{s(i)} = J_{r-m}(\beta^{(\gamma_i)})e^{i(r-m)\theta_{s(i)}^{(\gamma_i)}}, \label{eq:bessel}
 \end{equation}
 which depend on both the strength of the modulation index $\beta^{(\gamma_i)}$ and on the phase $\theta_{s(i)}^{(\gamma_i)}$ of the RF signal applied to the phase modulators in the measurement setting $\gamma_i$ \cite{lu2022bayesian,seshadri2022nonlocal}.
 After the transformation described by $U$, coincidences between the signal and the idler photon are collected in the bin combinations $(m,n)$, which are selected by using two fiber-Bragg-gratings (FBG). The center frequency of the stop band of the signal and the idler FBGs are set to be respectively $\omega_{n_s}$ and $\omega_{m_i}$. With good approximation, we can model the transfer function of the reflection port of the FBG (i.e., the one connected to the single photon detector) as a box 
  centered at $\omega_{n(m)_{s(i)}}$ and with a bandwidth of $2\Delta$. Since the frequencies within the stop band are not resolved, the action of the FBG is described by the operator $P_{mn}=P^{(i)}_m\otimes P^{(s)}_n$, where
 \begin{equation}
     P^{(i,s)}_q=\int_{-\Delta}^{\Delta} \ket{\omega_{q_{i(s)}}+\omega}\bra{\omega+\omega_{q_{i(s)}}}d\omega. \label{eq:S7}
 \end{equation}
Using Eqs.(\ref{eq:p_mn},\ref{eq:S7}) we obtain
\begin{equation}
    p_{mn} = \int_{-\Delta}^{\Delta}\int_{-\Delta}^{\Delta} \bra{\omega_{m_i}+\omega,\omega_{n_s}+\omega'}U\rho U^{\dagger}\ket{\omega_{n_s}+\omega',\omega_{m_i}+\omega} d\omega d\omega'.\label{eq:S8}
\end{equation}
We now use Eq.(\ref{eq:S6}) to write
\begin{equation}  U^{\dagger}\ket{\omega_{n_s}+\omega',\omega_{m_i}+\omega} = \sum_{\{r_i,q_s \}=-\infty}^{+\infty}(V_{rm}^{i}V_{qn}^{s})^*\ket{\omega_{q_s}+\omega',\omega_{r_i}+\omega} \label{eq:S9}
\end{equation}
which substituted in Eq.(\ref{eq:S8}) gives
\begin{align}
    p_{mn} = \sum_{\{ k,k'\}}\sum_{\{ r_i,q_s,r_i',q_s'\}} \int & \phi_{k}\phi_{k'}^{*}\braket{\omega_{r'_i}+\omega,\omega_{q'_s}+\omega'|\omega_{k_s}+\Omega_1,\omega_{k_i}+\Omega_2}\braket{\omega_{k'_i}+\Omega_2',\omega_{k'_s}+\Omega_1'|\omega_{q_s}+\omega',\omega_{r_i}+\omega}\times \\
    {} & \alpha_k\alpha_{k'}^* V_{mr'}^{i}V_{nq'}^{s}(V_{rm}^{i}V_{qn}^{s})^{*} d\omega d\omega' d\Omega_1 d\Omega_2 d\Omega_1' d\Omega_2' \label{eq:S10}
\end{align}
The different frequency bins do not overlap, and they are fully contained within bandwidth $2\Delta$ of the FBGs, which allows us to write
\begin{gather*}
\braket{\omega_{r'_i}+\omega,\omega_{q'_s}+\omega'|\omega_{k_s}+\Omega_1,\omega_{k_i}+\Omega_2}  = \delta(\omega'-\Omega_1)\delta(\omega-\Omega_2)\delta_{r_i',k_i}\delta_{q_s',k_s},  \\
\braket{\omega_{k'_i}+\Omega_2',\omega_{k'_s}+\Omega_1'|\omega_{q_s}+\omega',\omega_{r_i}+\omega}  = \delta(\omega'-\Omega_1')\delta(\omega-\Omega_2')\delta_{r_i,k'_i}\delta_{q_s,k'_s}. \label{eq:S11}
\end{gather*}
By performing the integrals over $\Omega_1,\Omega_2,\Omega_1',\Omega_2'$, and the sums over $\{r_i,r_i',q_s,q_s'\}$ we obtain
\begin{equation}
    p_{mn} = \sum_{\{k,k'\}}V_{mk_i}^{i}V_{nk_s}^{s}(V_{k'_i m}^{i}V_{k'_s n}^{s})^{*}\alpha_k\alpha_{k'}^*\int \phi_{k}(\omega_{k_s}+\omega',\omega_{k_i}+\omega)\phi_{k'}^{*}(\omega_{k'_s}+\omega',\omega_{k'_i}+\omega) d\omega d\omega'.\label{eq:S12}
\end{equation}
The overlap integral between the joint spectral amplitude $\phi_{k}$ and $\phi_{k'}$ on the right hand side of Eq.(\ref{eq:S12}) is the indistinguishability $I_{k k'}$ between source $k$ and $k'$, i.e.
\begin{equation}
    I_{k k'} = \int \phi_{k}(\omega_{k_s}+\omega',\omega_{k_i}+\omega)\phi_{k'}^{*}(\omega_{k'_s}+\omega',\omega_{k'_i}+\omega) d\omega d\omega'. \label{eq:S13}
\end{equation}
The final expression for the probability $p_{mn}$ of measuring a coincidence in the bin combination $(m,n)$ after that the transformation $U$ has been applied is then
\begin{equation}
    p_{mn} = \sum_{\{k,k'\}}\alpha_k\alpha_k'^*I_{k k'}V_{mk_i}^{i}V_{nk_s}^{s}(V_{k'_i m}^{i}V_{k'_s n}^{s})^{*} \label{eq:S14}
\end{equation}
We now demonstrate that the same probability outcome in Eq.(\ref{eq:S14}) can be obtained from a density matrix $\rho'$ with entries $\rho'_{ij,kl}=\bra{ij}\rho'\ket{kl}=I_{ik}\alpha_i\alpha_k^*\delta_{ij}\delta_{kl}$. The basis state $\ket{ij}$ describes a signal and an idler photon respectively in the frequency bins $\ket{i}$ and $\ket{j}$. Compared to the definition of $\rho$ in Eq.(\ref{eq:S5}), the density matrix $\rho'$ neglects any intra-bin frequency correlation in favor of a simpler description of the state, in which the frequency distribution is collapsed to a single label index. 
Note that in the limit $I_{ij}\rightarrow 1$, i.e. when all the sources are perfectly indistinguishable, we have that $\rho'\rightarrow \ket{\Phi}\bra{\Phi}$, where $\ket{\Phi}=\sum_{k=0}^{D-1} \alpha_k\ket{kk}$. When $|\alpha_k|=D^{-1/2}$, $\ket{\Phi}$ describes a maximally entangled Bell state. We can use Eq.(\ref{eq:p_mn}) to calculate the probability of measuring a coincidence over the bin combination $(m,n)$ after that the transformation $U$ has been applied, which gives
\begin{equation}
\begin{aligned}
    p_{mn}  & = \bra{mn}U\rho'U^{\dagger}\ket{mn} \\
    {} & = \sum_{i,j}\alpha_i \alpha_j^*I_{ij}\bra{mn}U\ket{ii}\bra{jj}U^{\dagger}\ket{mn}  \\
    {} & = \sum_{i,j}\sum_{p,q,r,s} (V^{i}_{pm}V^{s}_{qn})^*V^{i}_{mr}V^{s}_{ns}\alpha_i\alpha_j^* I_{ij}\braket{rs|ii}\braket{jj|pq}  \\
    {} & = \sum_{ij}\alpha_i\alpha_j^*I_{ij}(V^{i}_{jm}V^{s}_{jn})^*V^{i}_{mi}V^{s}_{ni},
\end{aligned}
 \label{eq:S15}
\end{equation}
which coincides with the expression in Eq.(\ref{eq:S14}). In the density matrix reconstruction algorithm described in section II, we neglect the intra-bin frequency correlations and represent $\rho$ on the $D^2$-dimensional basis $\ket{mn}$, where $\{(m,n)\,|\,m\in [0,..,D-1], n\in [0,D-1]\}$. By construction, the algorithm does not return the density matrix $\rho$ in Eq.(\ref{eq:S5}), but gives the maximum likelihood estimation of $\rho'$.   

\subsection{Impact of source indistinguishability on the purity and the fidelity of the reconstructed density matrix}

\begin{figure*}[h]
\centering
    \includegraphics[width=0.8\textwidth]{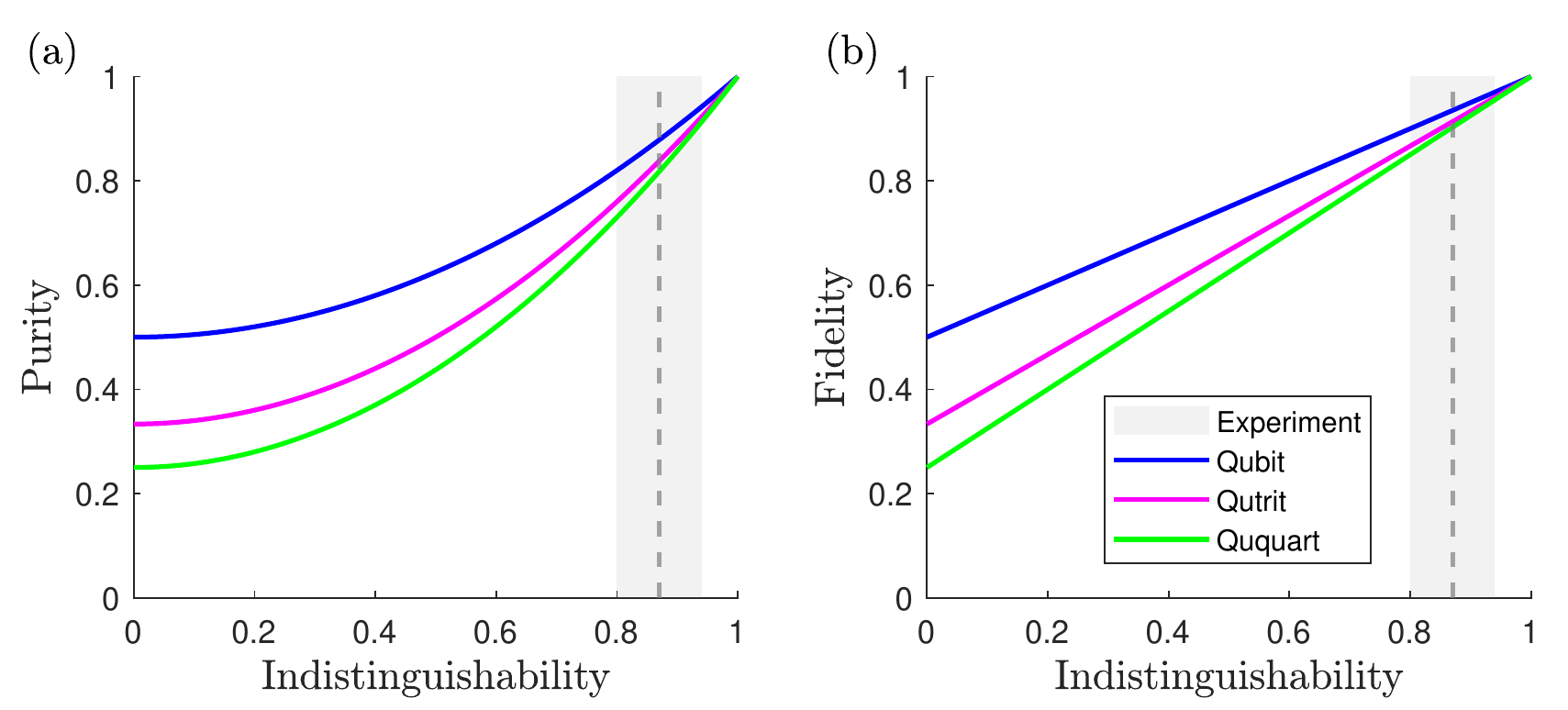}
    \caption{\label{fig:1} Simulated purity (a) and fidelity (b) with maximally entangled Bell states $\ket{\Phi}=\sum_{k=0}^{D-1}\ket{kk}$ of the density matrix $\rho'$ as a function of the indistinguishability $I$. The dashed line marks the average value of indistinguishability $\bar{I}=0.87(7)$ obtained in the experiment. The grey region represents the confidence bound of $\bar{I}$. }
\end{figure*}

The indistinguishability between all the sources determines the maximum amount of purity and fidelity with the target states that can be achieved by the reconstructed density matrices. 
We investigated how these quantities vary with the indistinguishability $I_{jk}$, which appear in the density matrix as $\rho'_{ij,kl}=\bra{ij}\rho'\ket{kl}=I_{ik}\alpha_i\alpha_k^*\delta_{ij}\delta_{kl}$. For simplicity, we assumed that all the sources have an equal pair-wise indistinguishability, i.e. $I_{jk}=I$.
Note that in the limit $I\rightarrow 1$, i.e. when all the sources are perfectly indistinguishable, we have that $\rho'\rightarrow \ket{\Phi}\bra{\Phi}$, where $\ket{\Phi}$ $=\sum_{k=0}^{D-1} \alpha_k\ket{kk}$. When $|\alpha_k|=D^{-1/2}$, $\ket{\Phi}$ describes a maximally entangled Bell state. 
In our analysis, we set $|\alpha_k|=D^{-1/2}$, and we calculated the purity $\mathcal{P}$ and the fidelity $\mathcal{F}$ with the maximally entangled Bell state $\ket{\Phi}$ as a function of $I$ for $D=2,3,4$. 
The results are shown in Fig.\ref{fig:1}. In the experiment, we measured an average  
 indistiguishability of $\bar{I}=0.87(7)$. This value is marked in Fig. \ref{fig:1} with a dashed line, while the gray area encloses the confidence bounds.
The range of values assumed by the  purity and the fidelity over that region are reported in Table \ref{tab:2}. These values are compared with the purities $\mathcal{P}$ and the fidelities $\mathcal{F}$ extracted from the reconstructed density matrices in the experiment. In Table \ref{tab:2}, we reported these values for the target states $\ket{\Phi_1}$ (qubit), $\ket{\Phi_2}$ (qutrit), and $\ket{\Phi_5}$ (ququart) (see definitions in Table 1 of the main text). Except for the purity of the qubit, which is slightly higher in the experiment than in the simulation, all the values lie in the confidence intervals predicted by the simulation. 

\begin{table}[h]
    \centering
    \begin{tabular}{c|c|c|c|c}
        \textbf{State} & \textbf{Simulated purity} & \textbf{Simulated fidelity} & \textbf{Experimental purity} & \textbf{Experimental fidelity} \\
        \hline \hline
        Qubit & $[81-94]$ & $[89-96]$ & $94.63(4)$ & $95.0(2)$ \\
        
        Qutrit & $[76-92]$ & $[86-96]$ & $78.3(5)$ & $86.2(3)$\\
        
        Ququart & $[73-91]$ & $[84-95]$ & $74.4(4)$ & $84.5(2)$\\
        
    \end{tabular}
    \caption{Measured and simulated purities and fidelities (percentage) for the qubit, qutrit and quaquart states. The interval bounds indicate the lower and the upper value of the purity and the fidelity over the confidence bounds of the indistinguishability $\bar{I}$. }
    \label{tab:2}
\end{table}

\newpage
%